\documentclass[a4paper]{article}

\usepackage{INTERSPEECH2022}
\usepackage{comment}
\usepackage{xcolor}
\usepackage{multirow}
\usepackage{tipa}

\title{Few-Shot Cross-Lingual TTS Using Transferable Phoneme Embedding}
\name{Wei-Ping Huang$^1$, Po-Chun Chen$^2$, Sung-Feng Huang$^1$, Hung-yi Lee $^1$}
\address{
 $^1$Graduate Institute of Communication Engineering, National Taiwan University \\
 $^2$College of Electrical Engineering and Computer Science, National Taiwan University
 }
\email{\{b07201054, b07901062, f06942045, hungyilee\}@ntu.edu.tw}

\begin{document}

\maketitle

\begin{abstract}
This paper studies a transferable phoneme embedding framework that aims to deal with the cross-lingual text-to-speech (TTS) problem under the few-shot setting. Transfer learning is a common approach when it comes to few-shot learning since training from scratch on few-shot training data is bound to overfit. Still, we find that the naive transfer learning approach fails to adapt to unseen languages under extremely few-shot settings, where less than 8 minutes of data is provided. We deal with the problem by proposing a framework that consists of a phoneme-based TTS model and a \textit{codebook} module to project phonemes from different languages into a learned latent space. Furthermore, by utilizing phoneme-level averaged self-supervised learned features, we effectively improve the quality of synthesized speeches. Experiments show that using 4 utterances, which is about 30 seconds of data, is enough to synthesize intelligible speech when adapting to an unseen language using our framework.
\end{abstract}


\section{Introduction}
Recently proposed TTS models based on deep learning techniques \cite{wang2017tacotron, shen2018natural, ping2018deep, sotelo2017char2wav} are capable of synthesizing natural, human-like speech. However, training end-to-end TTS systems requires large quantities of text-audio paired data and high-quality recordings; thus, data efficiency is a major challenge when developing advanced TTS systems for low-resource languages.


Transfer learning is a common approach when dealing with data efficiency problems. By training on rich-resource languages then fine-tuning on the unseen target language, the model benefits from cross-lingual information and improved data efficiency when adapting to the new language. A recent work from Microsoft \cite{yang2020towards} shows that multilingual training on 50 languages with a balanced data sampling strategy enables the model to adapt to a new language using only 6 minutes of paired data. Input linguistic features also play an important role in cross-lingual TTS. Many researches leverage common linguistic features across multiple languages to achieve cross-lingual knowledge-sharing. \cite{gutkin2017uniform, gutkin2017areal, demirsahinunified, maniati2021cross} use the combination of phonological features (PFs) and phonemes as inputs for multilingual neural TTS models and show improvements in intelligibility across both seen and unseen languages. \cite{chen2019end} utilizes International Phonetic Alphabet (IPA)\cite{international1999handbook} and an automatic speech recognition (ASR) system to build a unified symbol space and a cross-lingual phoneme mapping, respectively. Unicode-bytes-based models have also been proposed \cite{he2021multilingual, li2019bytes}, which create a unified input space by encoding input texts in UTF-8 and treating each byte as a token. Some works introduce techniques beyond basic transfer learning. For example, Lrspeech\cite{xu2020lrspeech} uses a high-resource language to bootstrap ASR and TTS systems for the low-resource language and leverages additional unpaired text and speech data to boost the performances of both systems.


While a variety of approaches have been proposed, methods that try to achieve cross-lingual knowledge-sharing through a unified input space still have some underlying issues. Using phonological features or IPA allows the low-resource language to leverage meaningful knowledge from rich-resource languages. However, these approaches require linguistic expertise, which can be quite unfeasible for dialects or minority languages. While using Unicode-bytes-based input methods bypasses the need for extra data and knowledge, transfer learning of phonological information cannot be achieved as bytes only encode typographical relations. To reduce the need for extra phonological knowledge, we propose learning a meaningful unified latent space for phonemes as a solution.

In this paper, we introduce few-shot cross-lingual adaptation using transferable phoneme embedding. We first pre-train a TTS model by leveraging data from high-resource (source) languages and then try to adapt it to low-resource (target) languages. To better transfer the cross-lingual knowledge from the source languages and tackle input space mismatch across languages, we propose a \textit{codebook} module as shown in Fig~\ref{fig:merge}b. The module utilizes speech representations extracted from self-supervised learned models to generate a better phoneme initialization for the target language. Compared to the baseline, our method generates more intelligible and natural speeches. Furthermore, experiments show that our method can produce intelligible speech on unseen language even with only 4 utterances, which is about 30 seconds of utterances from the low-resource target language. In comparison, the baseline requires more than 10 times the amount of data to generate intelligible speech.


\section{Few-Shot Learning}
\label{section:2}

Under the few-shot learning setting, we use \textit{tasks} to evaluate a model at test time. A \textit{task} contains $k$ training data and $q$ testing data. The model learns from the $k$ training data and is later evaluated on the $q$ testing data. We call this a $k$-shot task with $q$ queries. Our work deals with few-shot learning on cross-lingual TTS, where the testing tasks consist of data from unseen languages, and the goal is to learn a TTS model for the unseen language using only few-shot training data. We focus on learning one language at a time, meaning all data in a task belong to the same language.


\section{Method}
\begin{figure*}[t]
  \centering
  \includegraphics[width=0.8\linewidth]{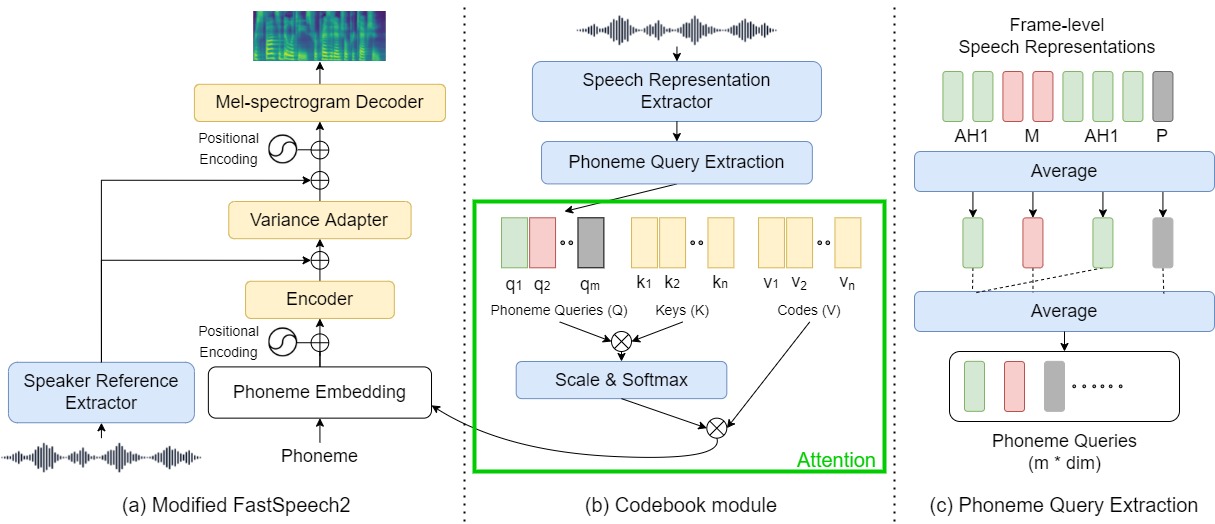}
  \caption{Illustration of the overall framework. (a) Modified multispeaker TTS architecture, (b) the codebook module and (c) the phoneme query extraction. The yellow blocks indicate trainable parts, while the blue blocks indicate fixed pre-trained models or non-trainable parts. $\mathbf{n}$: the size of codebook, $\mathbf{m}$: the size of phoneme set of each language, $\mathbf{dim}$: the dimension of speech representations, $\otimes$: matrix multiplication.}
  \label{fig:merge}
\end{figure*}

We propose a novel transfer learning framework for few-shot cross-lingual adaptation on TTS by utilizing a better phoneme embedding initialization. We illustrate the overall framework in Fig~\ref{fig:merge}. We first provide basic settings in Sec~\ref{section:3-1}, then introduce the baseline method in Sec~\ref{section:3-2}, and finally describe our proposed method in Sec~\ref{section:3-3}.

\subsection{Basic Settings}
\label{section:3-1}

We choose a modified version of FastSpeech 2\cite{ren2020fastspeech} from a previous paper\cite{huang2021meta} as our TTS architecture. However, the framework is not restricted to FastSpeech 2. For the multispeaker part, we extract speaker embeddings from a pre-trained speaker verification model. The speaker embeddings are then added to the inputs of both the variance adaptor and the decoder of FastSpeech 2. Fig~\ref{fig:merge}a shows the architecture of the modified FastSpeech 2. In addition to text-speech paired data, our method requires phoneme boundaries, which are already provided since they are needed for training FastSpeech 2. Note that under the few-shot setting, phoneme boundaries can still be manually aligned or forced aligned without too much labeling cost, so our framework can still be easily applied to other TTS architectures.

\subsection{Baseline: Multilingual Training}
\label{section:3-2}
For the baseline method, we train the TTS model on multiple languages together, then fine-tune the model on data from unseen languages. Since the TTS model is trained on multiple different languages, each with its own unique phoneme set, we use a randomly initialized language-dependent phoneme embedding table for each language. When testing, we also have to randomly initialize a phoneme embedding table for the unseen target language.


\subsection{Proposed: Phoneme Embedding Transfer}
\label{section:3-3}

When adapting to an unseen language using the baseline method, the encoder, the variance adaptor, and the decoder part of the model can be transferred. However, the embedding layer part has to be re-initialized, as mentioned in Section~\ref{section:3-2}. We introduce a novel \textit{codebook} module to project phonemes from different languages into a unified latent space and generate embedding tables. Our design is therefore capable of achieving phoneme-level knowledge transfer across multiple languages by sharing a common latent space.

\subsubsection{Phoneme Query Extraction}
To project phonemes from different languages into a unified latent space, we first extract \textit{phoneme queries} from audios as shown in Fig~\ref{fig:merge}c. For every distinct phoneme that appeared in an audio, we average frame-level speech representation vectors belonging to that phoneme according to its boundaries and take the result as a temporary representation of that phoneme. For multiple audios, we average all temporary representations of the same phonemes again to get the final \textit{phoneme queries}. For phonemes not appeared at all, we set them to zero vectors. Note that the process works regardless the phoneme has or hasn't been seen before. Since this is a general framework, it is possible to incorporate any kind of speech representation. Thus, we consider several different kinds of representations in this paper, such as Mel-spectrograms, or representations extracted from self-supervised learning (SSL) models such as hubert\cite{hsu2021hubert}, wav2vec 2.0\cite{baevski2020wav2vec}, and XLSR-53\cite{conneau2020unsupervised}.

\subsubsection{Codebook Module}
The \textit{codebook} module consists of phoneme query extraction and an attention module with learnable keys and values as shown in  Fig~\ref{fig:merge}b. We denote learnable keys and values as \textit{Keys} and \textit{Codes}. \textit{Phoneme queries} produced in the previous step are inputted as queries of the attention module. We apply scaled dot-product attention\cite{vaswani2017attention}. We expect \textit{Keys} to capture the patterns in phoneme queries, and \textit{Codes} to construct a unified latent space shared across different languages. \textit{Codes} can be viewed as a phoneme embedding basis of the latent space for all phonemes.


\subsubsection{Training and Fine-tuning}
In a single training step, every batch contains data from a single language. We then split the batch into two groups. Speech representations from one group are passed through the \textit{codebook} module to generate a phoneme embedding table. The training loss is calculated using the embedding table and the data from the other group.



When adapting to an unseen language, the only difference between our method and the baseline method is the phoneme embedding initialization stage. We generate a phoneme embedding table using the few-shot training data, and take the result as the initialization instead of randomly initializing it. The \textit{codebook} module is used to generate the phoneme embedding table at the beginning and is ignored at the fine-tuning stage.

\section{Dataset}

We use Mandarin, French, German, Japanese, and Korean in our experiments. Japanese and German are chosen for testing and the other 4 languages for training. We use LibriTTS\cite{zen2019libritts} train-clean-100 subset for English (\textit{en}), AISHELL-3\cite{shi2020aishell} for Mandarin (\textit{zh}), GlobalPhone\cite{schultz2002globalphone} for French (\textit{fr}), CSS10\cite{park2019css10} for German (\textit{de}) and Korean (\textit{ko}), and JSUT\cite{sonobe2017jsut} for Japanese (\textit{jp}). We use Montreal Forced Aligner\cite{mcauliffe2017montreal} to generate phoneme boundaries, and every language is independently aligned with their own phoneme set.

\section{Experiments}

\subsection{Training setup}
\label{section:5-1}
We split 10\% of the data of each language for validation. During the training stage, we extract 40 samples from the same language for each batch. 32 samples are used to generate the phoneme embedding table, while the remaining 8 samples are used to calculate the loss. We split the samples in such a way that any phoneme that appears in the remaining 8 samples also appears at least once in the 32 samples. This phoneme coverage condition ensures that embeddings used at the training step are meaningful since the \textit{phoneme queries} are only meaningful for phonemes that appeared in those 32 samples. The parameters are trained using Adam optimizer\cite{kingma2014adam} with an initial learning rate of 0.001 for 50K iterations. Warmup for 4K iterations and exponential learning rate decay is applied. For the \textit{codebook} module, the codebook size is set to 128, and we use multi-head attention\cite{vaswani2017attention} with 4 attention heads, each with 64-dim \textit{Codes}. The dimension of the embedding table of FastSpeech 2 is set to 256 to match the output from the \textit{codebook} module. We use MelGAN\cite{kumar2019melgan} vocoder from a pre-trained checkpoint released by the MelGAN’s authors to synthesize audios from Mel-spectrograms for all experiments.


\subsection{Few-Shot Language Adaptation}

For few-shot language adaptation, we test on two dissimilar languages: Japanese and German. We first find the minimum amount of data required for FastSpeech 2 baseline adaptation. We find that 64 shots, which is about 8 minutes of audio data, is the minimum amount of data required to fine-tune the multilingual baseline model to produce intelligible speech. Since we expect to push our model to the limit, we test our model under 4, 16, and 64-shot settings. The extreme 4-shot setting is very challenging since it contains only about 30 seconds of data. We compare models trained with features from the 24th layer of hubert-large, Mel-spectrogram features, and the baseline method. The ground truth audios are also evaluated for reference. Ground truth audios are first transformed into Mel-spectrograms and then resynthesized back to audios to show the influence of the vocoder. During testing, the speaker embeddings are extracted from the few-shot training data of each task.


We use Character Error Rate (CER) and Mean of Opinion Score test (MOS) as our evaluation metrics. Synthesized recordings are sent to Azure Speech-To-Text to get CER as a large-scale intelligibility metric. CER is averaged over 20 tasks to reduce the performance variance since only 30 seconds of training data is used in the most extreme case. Each task contains 64 queries. We construct the tasks in such a way that any phoneme that appears in the 64 queries also appears at least once in the few-shot training set. This way, we can ensure all phonemes used in the queries have a corresponding trained phoneme embedding. For the Mean of Opinion Score (MOS) test, we randomly sample 20 different sentences from all queries. Each sentence is rated by at least 5 individuals, and over 40 individuals are invited for both languages. The baseline method fails to produce intelligible speech under the 4-shot and 16-shot settings, so we exclude them from the MOS test. CER and MOS results are summarized in Table~\ref{tab:fse} and Table~\ref{tab:mos}. \textbf{Hubert} represents the model trained with features from the 24th layer of hubert-large, and \textbf{GT} represents ground truth audios resynthesized by the vocoder. 

Although our framework consistently achieves a higher naturalness score than other baselines, the overall score is low due to the few-shot setting. Some Japanese audios have been reported to sound like Korean, and some audios have unnatural intonation. Nevertheless, by using Mel-spectrogram features and the \textit{codebook} module, our method outperforms the baseline under few-shot settings, especially in the 4-shot and 16-shot categories, where the baseline model fails to produce intelligible speech. Furthermore, by using SSL features, the performance can be further improved. This suggests that transfer learning on phoneme embeddings improves the model's ability to generalize to different languages, and the model does indeed capture more shared knowledge between multiple languages.

\begin{table}[t]
\caption{CERs[\%] of recordings produced by different models under few-shot settings.}
\label{tab:fse}
\centering
\begin{tabular}{cccccc}
\toprule
\textbf{Lang.} & \textbf{Shots} & \textbf{GT} & \textbf{Baseline} & \textbf{Mel} & \textbf{Hubert} \\
\midrule
\multirow{3}{*}{\textit{jp}} & 4 & \multirow{3}{*}{16.52} & 81.27 & 44.51 & 25.26\\
& 16 & & 48.61 & 32.57 & 21.19 \\
& 64 & & 28.41 & 25.30 & 18.86 \\
\midrule
\multirow{3}{*}{\textit{de}} & 4 & \multirow{3}{*}{6.79} & 76.19 & 51.14 & 27.16 \\
& 16 & & 56.52 & 39.87 & 18.94 \\
& 64 & & 32.55 & 30.08 & 15.11 \\
\bottomrule

\end{tabular}
\end{table}

\begin{table}[t]
\caption{MOS of recordings produced by different models under few-shot settings.}
\label{tab:mos}
\centering
\begin{tabular}{cccccc}
\toprule
\textbf{Lang.} & \textbf{Shots} &  \textbf{GT} & \textbf{Baseline} & \textbf{Mel} & \textbf{Hubert} \\
\midrule
\multirow{3}{*}{\textit{jp}} & 4 &   & - & 2.21 & 2.80\\
& 16 & 4.32 & - & 2.79 & 3.01\\
& 64 &   & 2.97 & 3.28 & 3.42\\
\midrule
\multirow{3}{*}{\textit{de}} & 4 &   & - & 1.83 & 2.37\\
& 16 & 3.55 & - & 2.24 & 2.61\\
& 64 &   & 2.32 & 2.46 & 2.72\\
\bottomrule

\end{tabular}
\end{table}

\subsection{Feature Selection}

\begin{figure*}[t]
  \centering
  \includegraphics[width=0.88\linewidth]{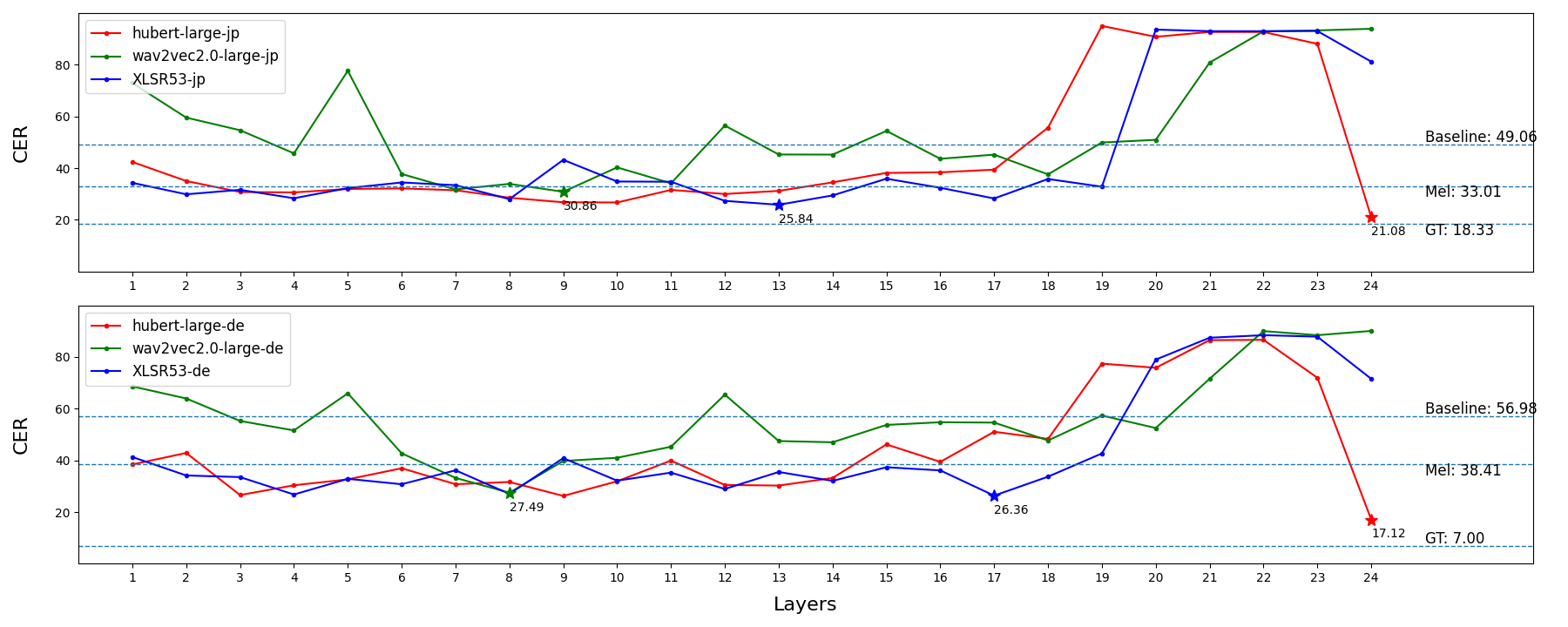}
  \caption{CERs of recordings produced by different feature extractors. Top: Japanese. Bottom: German.}
  \label{fig:20000-exp}
\end{figure*}

Since what we proposed is a general framework, it is possible to incorporate any kind of speech representation. Thus, we test our framework on several different extractors. We choose 4 different feature extractors: Mel-spectrogram converter, hubert-large, wav2vec 2.0-large, and XLSR-53. For SSL models with multiple layers, we test on all 24 layers of each model. Both hubert-large and XLSR-53 are SSL models built on the wav2vec 2.0 architecture. Hubert-large and wav2vec 2.0-large are both trained on English, but with different training objectives. XLSR-53 has the same training objective as wav2vec 2.0-large, but is trained on multiple languages.


Since there are dozens of models to evaluate, we only use CER as the evaluation metric. Each reported CER is averaged over 5 tasks, where each task contains 64 queries for CER testing. We use 16-shots for each task to show the transferability gap between the baseline method and our proposed framework since the baseline fails to synthesize intelligible speech under this setting. Results are summarized in Figure~\ref{fig:20000-exp}. We have also conducted the same experiment under the 64-shot setting. While the performance gap between the baseline method and our framework is narrower under the 64-shot setting, it still shows a similar trend as the 16-shot experiment. However, the result is not shown due to the space limit.


Most layers of SSL features outperform Mel-spectrogram features, and using Mel-spectrogram features is better than the baseline method. In general, XLSR-53 performs slightly better than wav2vec2.0-large, indicating that multilingual training is beneficial to multilingual tasks since XLSR-53 is built on wav2vec2.0 and trained on multiple languages. Hubert-large performs on par with XLSR-53 in most layers; however, its 24th layer shows an outstanding ability to extract features adaptable to different languages. We also observe that performances tend to heavily deteriorate at the last few layers for all three models, except the 24th layer of hubert-large. The exact reason is left unexplored.

Note that using features from some layers results in a worse error rate than the baseline method. We find that models using features extracted from those particular layers failed to learn the training languages during the training phase. From our experiments, we can see that the performance varies from layer to layer; therefore, layer selection should also be considered. Experiments show that under our proposed framework, features extracted from the 24th layer of hubert-large perform the best among all layers.

\subsection {Phoneme Mapping Discovery}

Inspired by \cite{chen2019end}, we study what the \textit{codebook} module has learned by analyzing its attention. First, we choose 256 sentences from every language in such a way that all phonemes appear at least once. Then, we generate the phoneme embedding table from them and extract the attention weights from the \textit{codebook} module. For any two distinct phonemes (which may come from different languages), cosine similarity is calculated between their attention weights from 4 attention heads and averaged as the final \textit{phoneme mapping score}. If two phonemes share similar attention weights among the attention heads, the score will be high, and vice versa.

For each phoneme, we list the top-5 phonemes from different languages with the highest \textit{phoneme mapping scores} and manually map them onto the International Phonetic Alphabets (IPA). Partial results of \textit{en} phonemes are shown in Table~\ref{tab:mapping}. Even in the mismatched cases, their corresponding IPA symbols still sound very alike. For example, \textit{ko}-ii, \textit{de}-I0, and \textit{jp}-i all sound like \textit{ee}, as in \textit{m\underline{ee}t}. There are also some failed cases, such as \textit{en}-TH. We find that the other five languages lack phonemes that correspond to '\texttheta' in IPA, and therefore don't share the same IPA as \textit{en}-TH. Intuitive mappings indicate that our framework can indeed learn a unified phoneme latent space successfully.


\begin{table}[t]
  \caption{Top-5 phoneme mapping discoveries of 10 en phonemes. A phoneme is marked blue if it shares the same IPA with the source phoneme.}
  \centering
  \begin{tabular}{ccc}
    \toprule
    \textbf{Source} & \textbf{IPA}     & \textbf{Top-5 closest phonemes} \\
    \midrule
    \textit{en}-AA0 & \textscripta &\textcolor{blue}{\textit{de}-a1}, \textit{de}-\&1, \textcolor{blue}{\textit{zh}-a1}, \textcolor{blue}{\textit{zh}-a4}, \textit{de}-O0\\
    \textit{en}-AO1 & \textopeno &\textcolor{blue}{\textit{de}-O0}, \textcolor{blue}{\textit{de}-O1}, \textit{jp}-w, \textit{ko}-wv, \textcolor{blue}{\textit{fr}-O}\\
    \textit{en}-CH & \textteshlig &\textcolor{blue}{\textit{de}-J}, \textit{zh}-ch, \textit{zh}-c, \textit{zh}-q, \textit{ko}-ch\\
    \textit{en}-D & d &\textcolor{blue}{\textit{de}-d}, \textit{de}-t, \textcolor{blue}{\textit{jp}-d}, \textcolor{blue}{\textit{ko}-t0}, \textit{ko}-c0\\
    \textit{en}-EY2 & e\textsci &\textit{ko}-we, \textcolor{blue}{\textit{zh}-ei3}, \textcolor{blue}{\textit{zh}-ei2}, \textit{de}-e0, \textit{jp}-e\\
    \textit{en}-M & m &\textcolor{blue}{\textit{de}-m}, \textcolor{blue}{\textit{jp}-m}, \textcolor{blue}{\textit{fr}-M}, \textcolor{blue}{\textit{zh}-m}, \textcolor{blue}{\textit{ko}-mm}\\
    \textit{en}-OW2 & o\textupsilon &\textcolor{blue}{\textit{zh}-ou2}, \textcolor{blue}{\textit{zh}-ou3}, \textit{de}-U0, \textit{de}-U1, \textcolor{blue}{\textit{zh}-ou4}\\
    \textit{en}-S & s &\textit{jp}-sh, \textcolor{blue}{\textit{de}-s}, \textit{ko}-ss, \textcolor{blue}{\textit{zh}-s}, \textcolor{blue}{\textit{fr}-S}\\
    \textit{en}-TH & \texttheta &\textit{jp}-z, \textit{zh}-z, \textit{zh}-f, \textit{de}-f, \textit{jp}-u\\
    \textit{en}-Y & j &\textcolor{blue}{\textit{de}-j}, \textit{jp}-ry, \textit{ko}-ii, \textit{de}-I0, \textit{jp}-i\\
    \bottomrule
  \end{tabular}
  \label{tab:mapping}
\end{table}

\section{Conclusions}

In this paper, we propose a framework for few-shot cross-lingual adaptation on text-to-speech. By learning a unified phoneme latent space using SSL representations, the model can adapt to an unseen language with only 30 seconds of data. Different experiment settings show that our method outperforms the naive transfer learning baseline by a large margin in terms of performance and data efficiency.

\section{Acknowledgements}
We thank to National Center for High-performance Computing (NCHC) of National Applied Research
Laboratories (NARLabs) in Taiwan for providing computational and storage resources.

\bibliographystyle{IEEEtran}
\bibliography{mybib}

\end{document}